# Bio-Inspired Synergistic Wing and Tail Morphing Extends Flight Capabilities of Drones

E. Ajanic[a*], M. Feroskhan[b], S. Mintchev[c], F.Noca[d], D. Floreano[a*]

[a]Laboratory of Intelligent Systems, École Polytechnique Fédérale de Lausanne, Switzerland; [b]School of Mechanical and Aerospace Engineering, Nanyang Technological University, Singapore, Singapore; [c]Reconfigurable Robotics Laboratory, École Polytechnique Fédérale de Lausanne, Switzerland;  [d]Haute école du paysage, d'ingénierie et d'architecture de Genève, Switzerland.

Correspondence to `{enrico.ajanic, dario.floreano}@epfl.ch`

**Abstract -** The operation of drones in cluttered environments and over extended areas demands adaptive flight capabilities to meet the opposing aerodynamic requirements of agile and fast cruise flight. High agility and maneuverability are required to aggressively navigate around obstacles and to perform instantaneous takeoffs or landings, while high energy efficiency is desired when covering large distances. In nature, these requirements are met by some birds by synergistic adaptation of wings and tail, such as the northern goshawk, which displays high agility and maneuverability when flying through forests and fast steady flight capabilities when ambushing prey in the open field. In this article, we experimentally study the effects of bio-inspired wing and tail morphing on flight performance by means of a novel morphing drone. We show that the combined morphing of wing and tail can improve agility, maneuverability, stability, flight velocity range, and energy efficiency of a winged drone. The drone's flight performance is validated in wind tunnel tests, shape optimization studies and outdoor flight tests.

## Introduction

Fixed-wing drones play an increasing role in civilian applications, such as disaster mitigation, environmental monitoring, inspection, and delivery, to mention a few (*1*). Their aerodynamic efficiency enables fast cruise flight for covering larger distances with lower energy expenditure than multicopters of the same mass. However, fixed-wing drones still struggle when navigating in complex, obstacle rich environments, such as cities (*2*, *3*).

For aggressive flight in complex environments, fixed-wing drones must perform sudden and sharp course variations at a broad velocity range, requiring agility, maneuverability and a low inherent stability. *Agility* and *maneuverability* are here defined as the abilities of inducing high angular and linear accelerations, respectively (Materials and Methods). High agility allows the drone to rapidly change its body orientation in space (*4*). The agility is greatest when the turning moments produced by lifting surfaces' geometry (wing and tail) are large, the airframe's moment of inertia is small, and the *inherent stability* is low (*5*, *6*). High maneuverability ensures that the drone can promptly alter its flight trajectory around obstacles (*7*, *8*). Thus, the maneuverability of a drone is greatest when a strong propulsive force is applied or the aerodynamic forces such as lift and drag produced by the lifting surface's geometry are large. Instead, the aerodynamic requirements for fast cruise flight are very different. Here, the lifting surfaces should be small in order to decrease parasitic drag and to reduce sensitivity to head winds (*9*). Contrary to aggressive flight, in fast-cruise flight high inherent stability is favored because it helps the drone maintaining equilibrium flight condition, thus being less sensitive to wind gusts (*10*). A fixed-wing drone cannot effectively resolve these opposing dynamic requirements because the design of its lifting surfaces excels only within a small range of operating conditions (*11*, *12*).

Birds overcome this problem by synergistic morphing of wing and tail to increase their flight performance when gliding (*13*). To increase agility, birds enlarge and deflect their wings and tails to produce considerable aerodynamic moments (Fig. 1A left) (*14*). The relatively short body length of birds, which resembles more a flying wing than a traditional aircraft, and their lightweight wings contribute to lower inherent inertia, which further enhances agility (*15–17*). Birds sweep their main wing forward and fan the tail outward to produce considerable aerodynamic forces such as lift and drag during slow and aggressive flight (Fig. 1A, left) (*18*). While the avian tail is used to increase agility, research suggests that its principle role appears to supporting the main wing in generating lift and drag (and thus improving maneuverability), analogous to trailing edge flaps on aircraft wings (*19*, *20*). Furthermore, birds actively decrease their longitudinal stability by sweeping the main wings forward and minimizing the tail's surface (*21*, *22*). Instead, during fast cruise flight (Fig. 1A, right), birds reduce the size of wing and tail and sweep their wings backward to generate a streamlined profile that reduces drag forces (*23–25*). This backward swept wing configuration also increases the inherent longitudinal stability (*13*).

Bird's aerodynamic adaptability has inspired researchers to investigate avian-inspired morphing strategies for drones (*26*). Variable sweep/area (*9, 27–31*), variable dihedral (*32, 33*), and variable twisting (*34, 35*) wing morphing have shown to improve aerodynamic performance and to extend mission capabilities. However, the focus of these experimental studies has been limited to the main wing. The synergistic role of morphing both wing and tail, as observed in birds, has not yet been systematically studied.

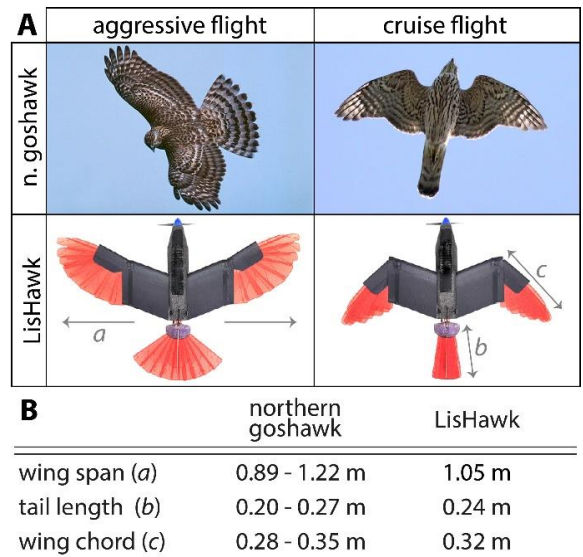

Fig. 1. **Avian-inspired synergistic morphing increases flight performance during both aggressive and cruise flight.** (**A**) Two examples of avian morphing seen on the northern goshawk compared to the LisHawk drone. During aggressive flight, both sweep their wing forward and increases its tail area (left). During fast cruise flight, their wing is swept backward and their tail area is decreased to reduce parasitic drag (right). Photo credit: (36, 37). (**B**) The LisHawk's size and wing/tail proportions are inspired by the northern goshawk (accipiter gentilis) (38).

Here, we experimentally study the aerodynamic benefits of avian-inspired, synergistic morphing of tail and wing (Fig. 1A right) for increasing a drone's mission capability by adapting the aerodynamic profile to different and contrasting flight regimes, namely aggressive flight and cruise flight. We adopt a wing and tail design based on artificial feathers (11) and implement it on a drone (codenamed LisHawk) to synergistically adapt the sweep angle and area of its lifting surfaces during flight. The drone dimensions approximate those of the northern goshawk (*Accipiter gentilis*), which is a migratory bird capable of both, quick maneuvering in highly cluttered environments, such as forests, as well as fast gliding flight when hunting in the open terrain (*38, 39*). The experimental method consists in the aerodynamic characterization, shape optimization, and flight testing of the avian-inspired drone in different morphing configurations. Specifically, we want to show that the synergistic application of a morphing wing and tail can improve agility, maneuverability, inherent stability, and energy efficiency when changing between tucked wing and tail (cruise flight) and extended wing and tail (aggressive flight) configuration.

## Results

**LisHawk drone.** Our aim was to understand how synergistic morphing as seen on the gliding northern goshawk could extend flight capabilities of winged drones and how such morphing surfaces could be implemented into a drone. Therefore, we developed the LisHawk drone composed of a morphing main wing and a morphing tail (Fig. 1A) with size and proportions inspired by the northern goshawk (Fig. 1B). The northern goshawk's lifting surfaces consist of a relatively long, wedge-tipped tail and short, broad wings with a large wing chord (Fig. 1A and B) that enable large geometrical changes to efficiently perform both aggressive maneuvers when hunting in dense forest and fast gliding flight when silently approaching its pray in the open field (*38, 40–42*). For the morphing wing and tail we implemented artificial feathers (Fig. 2E), which are similar to the design introduced by (*9*). While real northern goshawk feathers could be advantageous due to their low weight, their softness, and their self-healing properties, the geometry variation of individual feathers between different birds can be large (*30*), which hampers scalability and the ease of drone production. Our artificial feathers (Fig. 2E), however, are adaptable to a wide variety of designs, offer repeatability as they can be precisely manufactured, and are durable because of their soft architecture. Furthermore, the northern goshawk's body structure is comprised of rigid, lightweight bones that absorb loads and aerodynamic surfaces made from soft flesh and feathers.

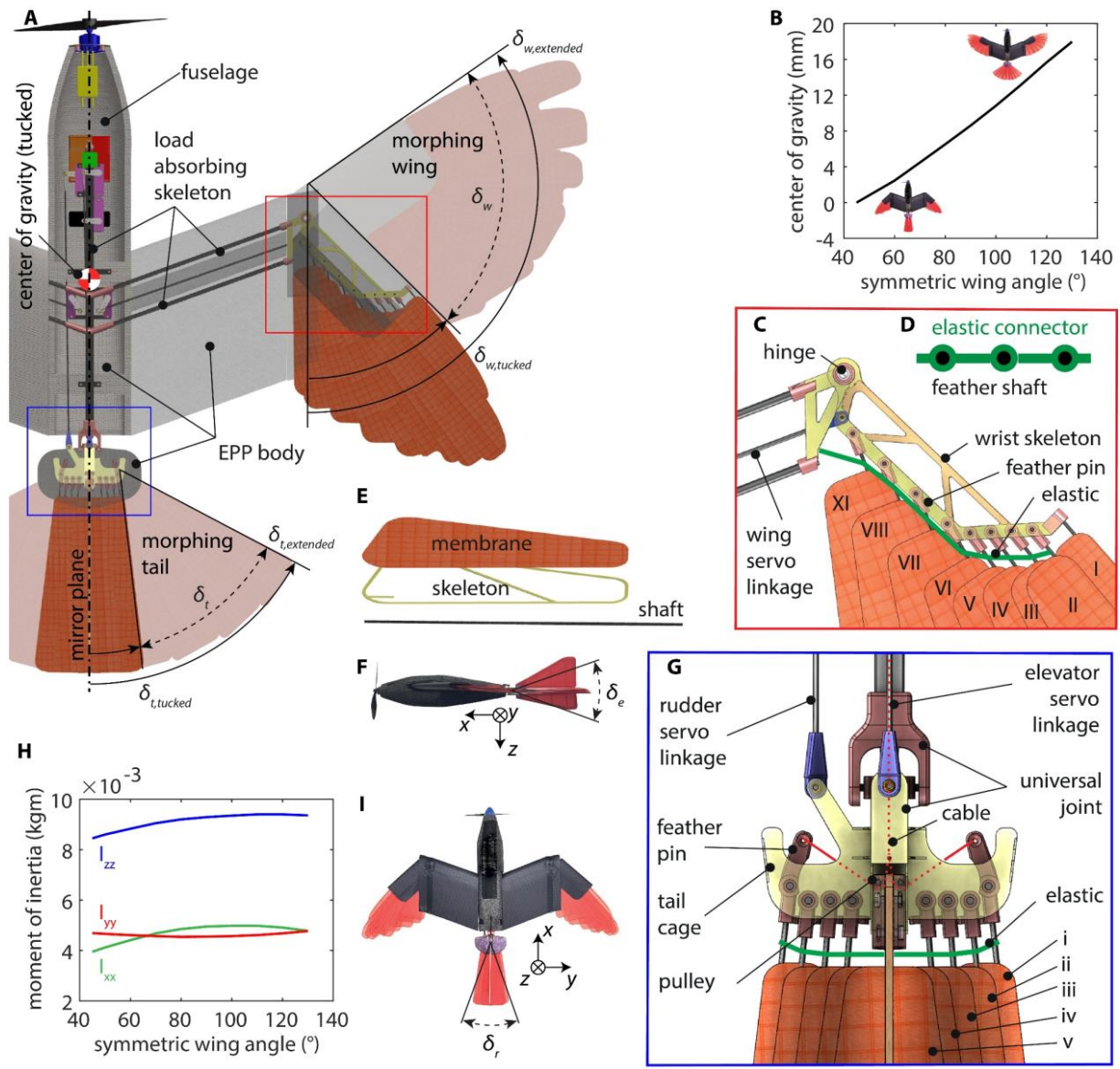

Fig. 2. **LisHawk morphing platform architecture**. (**A**) Top view of the LisHawk MAV with a partially transparent body to show the load absorbing skeleton, the morphing angles of wing and tail, and the location of the drone control system. The drone control system (see Materials and Methods for more details) consist of the motor (blue), the electronic control board and power module (yellow), the autopilot (red), the receiver (orange), the global positioning system (green), the battery (black), and the five servomotors (purple). The top two servomotors actuate the rudder (left) and the elevator (right), the middle servomotor actuates the tail spread, and the bottom two servomotors actuate the wing sweeping on their corresponding sides. The main wing can continuously sweep from tucked ($\delta_{w,tucked} = 45°$) to extended ($\delta_{w,extended} = 130°$) over a range $\delta_w = 85°$. The tail can change its sweep from tucked tail $\delta_{t,tucked}$ to a extended tail $\delta_{t,extended}$ over a range $\delta_t = 50°$ (Mov. S3). (**B**) Nearly linear change in center of gravity by 18.6 mm due to symmetric wing morphing (CAD data). The tail's contribution when morphing is negligible. (**C**) Top view of the right wing morphing mechanism (see Materials and Methods for further information, Mov. S1). For sake of clarity, the EPP body is removed. (**D**) A 3D-printed elastic connects the feather shafts. (**E**) The artificial feathers consist of three parts: a durable ripstop membrane, a flexible glass fiber skeleton and a stiff carbon shaft. (**F**) The horizontal tail can deflect in the vertical plane ($\delta_e = \pm 20°$) to act as an elevator. (**G**) Top view of the tail morphing mechanism (see Materials and Methods for further information). A universal joint is used to allow elevator and rudder deflections. The feathers are held by a tail cage through pins. For sake of clarity, the body is removed. (**H**) Wing morphing from tucked to extended changes the moment of inertia by 26% for $I_{xx}$, by 5% for $I_{yy}$, and 11% for $I_{zz}$ (CAD data). The other moment of inertia components and the tail's contribution when morphing as small and thus not shown (**I**) The vertical tail can deflect in the horizontal plane ($\delta_r = \pm 20°$) by $\delta_r$ to act as a stabilizing surface and a rudder $\delta_r$.

As a result, their overall mass is low and their wing's moment of inertia is small (*20*), which increases maneuverability and agility, respectively (see Materials and Methods). Similarly, to keep the drone's weight and moment of inertia low while providing sufficient mechanical robustness, we designed a skeleton made from fiber reinforced plastics to provide a high strength and load absorption at a low weight, which is encapsulated by a durable, flexible, and lightweight expanded polypropylene (EPP) body (Fig. 2A). We also placed the drone control system (52 % of the overall drone weight) close to the drone's center of gravity (Fig. 2A) to further reduce the moment of inertia. Unlike the goshawk, which generates thrust by flapping its wings, we implemented a tractor propulsion system consisting of an electrical motor and a propeller. This offers a high propulsive efficiency at a low system complexity (*43*). All these design strategies lead to a drone with a ready-to-fly mass of 284 g with a flight time of 10 minutes.

**Morphing wing and tail architecture.** Birds such as the northern goshawk greatly vary the area of their main wing through wrist folding (*44*). As such, we applied a variable sweep and area morphing mechanism on the outer sections of both wing sides, while the inner section remains fixed (Fig. 2A). The morphing wing consists of nine artificial feathers (I-IX), which fan outward when the wing is extended and overlap on each other when the wing is tucked (Fig. 2C). The outermost feather (I) is fixed to the skeleton, while the inner feathers (II to IX) can rotate in the wing plane around their feather pins. To actuate feathers II to IX, we implemented an elastic connector between each feather (Fig. 2C) to achieve regular feather spacing in the wing plane when the wing is extended (*9*). This connector is pre-stretched, which helps to overcome the surface friction of the overlapping feathers when the wing is tucked. This avian-inspired, under actuated design strategy was chosen, because it is lightweight and it increases the mechanical robustness of the morphing surfaces due to its softness (*30*). Each side of the wing is independently actuated by a separate servo motor to continuously adjust the sweep angle (Fig. 2A, Mov. S3). The synchronous activation of the two wing sides generates a symmetric sweep variation, resulting in a maximum wing area change of 41 % (Tab. 1). This change in wing morphology also shifts the center of gravity forward (Fig. 2C) and changes the drone's moment of inertia (Fig. 2H). Asynchronous activation generates asymmetric sweeping between the two wing sides, resulting in a maximum area difference between left and right wing of 40 %. This area divergence produces a moment around the aircraft's x-axis (Fig. 2F), which is used to control roll. We chose asymmetric morphing as opposed to ailerons or wing twisting because previous studies suggested asymmetric sweeping produces greater moments in the high angle of attack regime than ailerons, which could increase agility during aggressive flight (*9*, *45*).

Tab. 1. **Geometrical properties of the LisHawk's morphing wing and tail.**

| property | measurement | |
| --- | --- | --- |
| | tucked | extended |
| wing sweep (deg) | 45 | 130 |
| wing area (m$^2$) | 0.117 | 0.165 |
| tail sweep (deg) | 10 | 60 |
| tail area (m$^2$) | 0.014 | 0.044 |
| wing & tail area (m$^2$) | 0.131 | 0.209 |

The goshawk's tail can fold its feathers to change its area, deflect upward/downward to act as both an elevator and a flap, as well as twist to act as a control device in yaw (*14*). Similarly, we developed a feathered tail which can change its area and can deflect upward/downward (Fig. 2F, Mov. S3), yet we forwent the tail twisting. Instead, to simplify the design, the tail can deflect sideways (Fig. 2I) to induce a yaw moment through a double vertical fin (Mov. S3). Our morphing tail consists of 9 artificial feathers - one fixed central feather (v) and four feathers on each side of the central feather (i to iv), which can rotate around the feather pins (Fig. 2G). Both sides of the tail are actuated in symmetry. Analogous to the morphing wing, we interconnected the tail feathers by a pre-stretched elastic connector for even spacing in the tail plane, while the outermost feather (i) guides the inner feathers (ii to iv) when extending the tail (Fig 2G). The outer feather (i) is actuated via cable that is guided over pulleys to a servo at the front of the LisHawk's fuselage (Fig. 2A). The tail extends when the cable is pulled, while the elastic connector tucks the feathers, when the servo tension is released. This allows an area change from fully tucked to fully extended of 214.3 % (Tab. 1).

**Synergistic morphing increases maneuverability.** We next examined the increase in maneuverability as a result of wing and tail morphing. Maneuverability describes the drone's controlled ability to change its velocity vector (*7*), which is greatest when the linear accelerations acting on the airframe are maximized. These linear accelerations are dependent on the aerodynamic forces such as lift, drag (Fig. 3A), and the weight force when excluding the thrust force (see Materials and Methods for mathematical formulation). Increased lift permits a fast change in flight path direction, which is essential when navigating complex environments or avoiding obstacles, while increased drag decelerates the drone faster, which is required for maneuvers such as perching (*46*). Furthermore, the weight force should be small which was considered during the drone's design process.

By placing the LisHawk drone in the wind tunnel (see Material and Methods), we assess the change in lift and drag forces due to morphing (Fig. 3B to D). Our experiments show that changing the wing and tail from tucked to extended (59.9% area increase) increases lift (70.8%) and drag (63.8%) (Fig. 3D). As commonly observed in the low Reynolds number regime ($< 10^5$) (*20*, *41*), the maximum lift is reached at high angles of attack (approx. 24 degrees) for all configurations, and remains nearly constant beyond that angle, while drag further increases. When measuring the wing's independent contribution to the aerodynamic forces (Fig. 3C), we find that wing morphing steepens the lift slope, which increases lift at low angles of attack (for example 30 % at 5 degrees) and at high angles of attack (maximum lift: 49.6 %). Similarly, drag is increased over the entire measured angle of attack range, which can benefit maneuverability also at low angles of attack. Enlarging the tail (22.9% overall area change), however, only increases lift (20.1%) and drag (14.7%) in the high angle of attack regime, while at low angles of attack there is no clear increase identifiable (Fig. 3B). We think that this could be due to the disturbed flow conditions behind the fuselage's and the wing's wake, which directly affect the adjacent morphing tail (*16*). Finally, we apply our static wind tunnel measurements to our maneuverability metric (Materials and Methods). We can show, that at the measured flight velocity of 8 m/s, for example, when changing the flight path direction maneuverability increases by 175.8 %, while maneuverability when decelerating increases by 137.6 %.

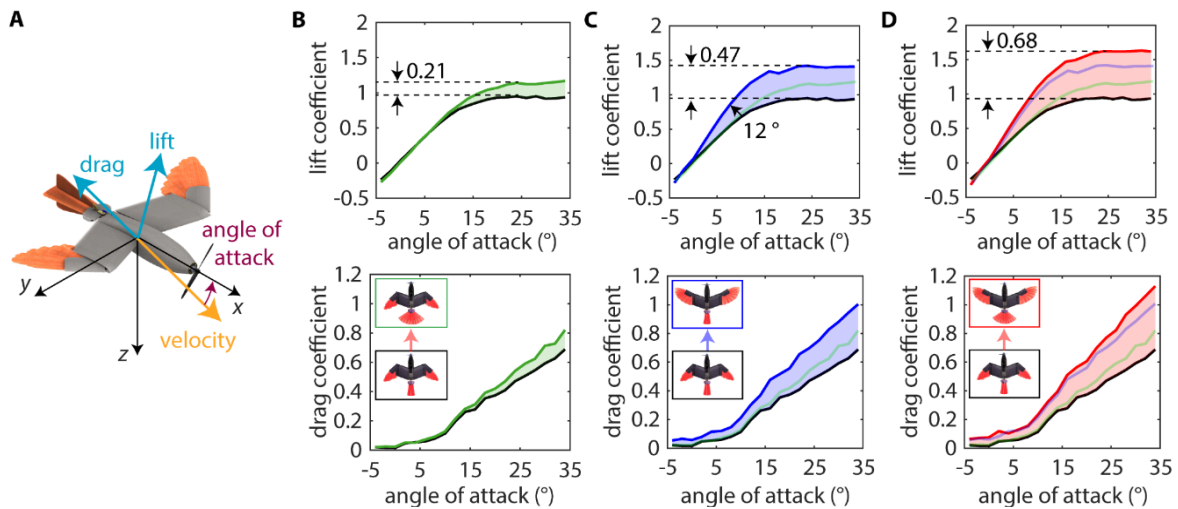

Fig 3. **Avian-inspired morphing of wing and tail improves maneuverability by increasing lift and drag.** Wind tunnel data (flow velocity of 8.0 m/s) with the LisHawk when changing from the cruise flight configuration (black) to large tail (green), large wing (blue), and both large wing and tail (red) at different angles of attack. The shaded areas indicate the lift and drag coefficients for intermediate morphing configurations. The dashed black lines indicate the location of the maximum lift. (**A**) We presume the drag force parallel to the velocity vector (the velocity vector deviates from the x-axis by the angle of attack in the x-z plane), while lift is perpendicular to the velocity vector and the y axis. (**B**) When extending the tail (wing tucked) lift and drag coefficients only increase at higher angles of attack beyond 14°. (**C**) Extending the wing increases lift at low angles of attack due to a change in lift slope. (**D**) The synergistic enlargement of wing and tail increases lift and drag for all positive measured angles of attack.

**Synergistic morphing increases agility.** We now proceed with the study of the drone's controlled ability to change the angular rate, which we term agility. It is greatest when the angular accelerations acting on the airframe are maximized, which is the case if the aerodynamic moments (pitch, roll, yaw) (Fig. 4E) are large and the moments of inertia small (see Materials and Methods for mathematical formulation). Specifically, we focus on the positive (nose up) pitch moment and the roll moment, which are key factors during aggressive flight when performing maneuvers, such as perching or rapid turning (*14*, *47*). Furthermore, it is important to note that agility and maneuverability are closely linked. A high positive acceleration in pitch temporarily increases the angle of attack (*5*), which in return increases maneuverability (Fig. 3B).

First, we ask to what extend synergistic morphing could increase the nose up pitch moment as a result of tail deflection and wing and tail morphing (Fig. 4A to D). Our results show that avian inspired morphing provides a notably larger pitch moment over the measured angle of attack range, compared to the non-morphing drone (Fig. 4D). As can be observed when extending the tail (Fig. 4A), the nose up pitch moment is increases by 253.9 % (at 4° incidence), which suggests a substantial increase in agility (Fig. 3B). Moreover, the transition point from the positive to negative pitch moment is delayed from an angle of attack of 11° to 17°. Indeed, when the wing is tucked, agility can be limited to the low angles of attack regime, which is due to the drone's inherent stability discussed in the next section. This limitation is overcome by extending the wing, which has previously shown to shifts the overall center of lift forward relative to the center of gravity (Fig. 2B), thus producing a positive pitch moment (*13*, *27*). In doing so, our measurements indicate that the nose up pitch moment increases even at high angles of attack (Fig. 4C and D). Beyond 20° angle of attack, the tail produces positive lift which induces a negative moment. This reduces the overall nose up pitch moment. Hence, tucking the tail while the wing is extended is the preferred morphology at angles of attack >20° as compared to the extended wing and tail, as it reduces the negative moment induced by the tail. Therefore, the wing and tail morphology to induce a large pitch moment is dependent on the respective flight regime. For example at low angles of attack (<-2°) the wing should be tucked and the tail extended, at median angles of attack (-2…20°) the wing and tail should be extended, and at high angles of attack (>20°) the wing should be extended and the tail tucked (Fig. 4D). When applying our findings to the agility metric (Materials and Methods), we can show, for example, at an angle of attack of 4° a pitch agility increase of 239.6% as compared to the fully tucked configuration. Moreover, even at high angles of attack (for example at 30°) when extending the wing and tucking the tail, we surpass the maximum agility of the fully tucked wing and tail configuration by 59.9%.

We also studied the roll moment as a result of asymmetric wing sweeping. Here, we extend previous findings (*9*, *30*) to explore its performance in the high angle of attack regime. Our results confirm that there is no decrease in roll moment at high angle of attack, suggesting good roll behavior during aggressive flight (Fig. 4F and G). Since the roll moment for the asymmetric morphing wing is proportional to the wing's lift force, this is in accordance with the nearly constant lift at high angles of attack, seen in Fig. 3B.

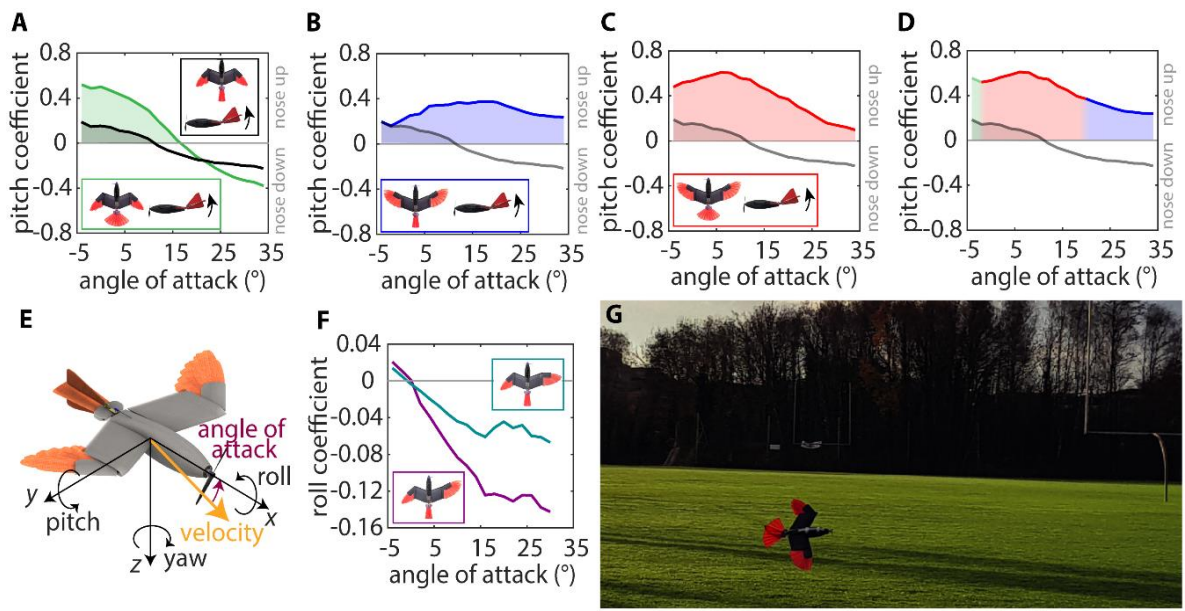

Fig. 4. **Avian-inspired morphing of wing and tail improves agility by increasing the pitch and roll moment.** Wind tunnel data (flow velocity of 8.0 m/s) with the LisHawk when deflecting the elevator upward by 20° changing from fully tucked wing and tail (black) to large tail (green), large wing (blue), and both large wing and tail (red), as well as when applying asymmetric sweep at different angles of attack. The shaded areas indicate the change in positive pitch moment with respect to the zero moment (gray line). (**A**) The relatively low pitch coefficient when wing and tail are tucked can be increased when extending the tail. However, a positive pitch moment can only be produced at low angles of attack. (**B**) Extending the tail increases the pitch coefficient at high angles of attack. (**C**) Extending wing and tail notably increases the pitch coefficient over the entire angle of attack range. (**D**) The greatest positive pitch moments can be produced when switching between different morphologies with respect to the angle of attack. (**E**) Moment coefficients are reported in the body frame. (**F**) Both wing asymmetries of 45° (teal) and 80° (purple) show a high roll coefficient at high angles of attack. (**G**) Full asymmetric sweep applied during flight tests for roll control (Mov. S4).

**Synergistic morphing changes pitch stability.** To understand how the morphing of wing and tail might affect the pitch stability, and thus flight performance, we studied the gradient of the pitch coefficient curves for different wing and tail morphologies (Fig. 5B to D). A negative gradient exhibits a stable aircraft, while a positive gradient implies an unstable aircraft (mathematical formulation in Materials and Methods). On one hand, stability in pitch is desirable during fast cruise flight as the aircraft returns to trim (pitch moment is zero and the aircraft is in equilibrium) when disturbed and it can be controlled with ease (*5*). On the contrary, low pitch stability or instability can be desirable during aggressive flight because light control forces can produce large rotational accelerations. However, such aircraft are often difficult to control and must be actively stabilized by a sophisticated control system (*41*).

In comparison to previous studies of bird gliding flight (*13*), we can confirm that the morphing of wing and tail independently or in synergy also allows to greatly adapt stability and the trim. While we designed the LisHawk to be stable in pitch by placing the drone's center of gravity accordingly, we can show that extending the tail further increases pitch stability by decreases the gradient of the pitch curve (Fig. 5B). We explain this by the increased lift and drag (Fig. 3C) produced behind the drone's center of gravity, which increases the restoring moment and thus stability (see Supplementary Material for mathematical formulation). When extending the wing, our data show a positive gradient at low angles of attack, which suggests instability (Fig. 5C). This is caused by the shift in lift and drag when extending the wing, which is greater in magnitude than the change in center of gravity (Fig. 2C). However, at high angles of attack the gradient transitions into the negative, suggesting a stable trim point at > 34° when extending the wing and tucking the tail (Fig. 5B) and at 20° when extending wing and tail (Fig. 5C). Furthermore, changing from tucked wing and tail to extended wing and tail shifts the trim point incidence from 4° to 20° while remaining inherently stable, which increases lift by 328% and drag by 580%.

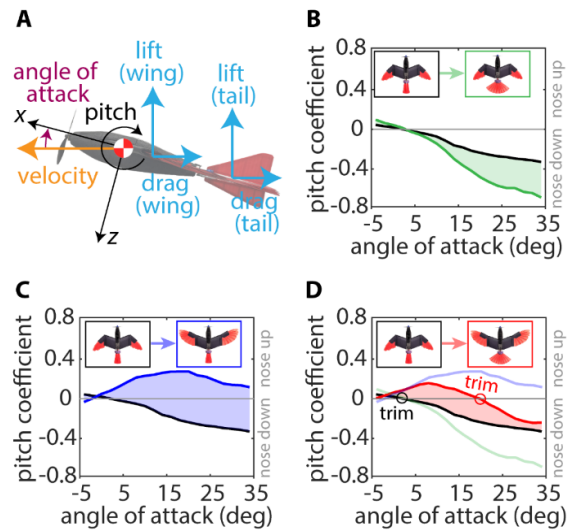

Fig. 5. **Synergistic morphing greatly alters pitch stability which improves agility.** Wind tunnel data (flow velocity of 8.0 m/s) with the LisHawk when changing from fully tucked wing and tail (black) to tucked wing and extended tail (green), extended wing and tucked tail (blue), and extended wing and tail (red) at different angles of attack. The shaded areas indicate the coefficients for intermediate morphing configurations. (**A**) Lift and drag produced by wing and tail induce a pitch moment around the aircraft's center of gravity. For aircraft that are stable in pitch, these forces produce a negative moment at positive angles of attack and a negative moment at negative angles of attack. (**B**) Extending the tail increases the pitch stability, by increasing the negative (nose down) pitch moment at high angle of attack. (**C**) Extending the wing leads to instability at low angles of attack up to an incidence of 20°. (**D**) Extending the wing and tail in synergy shifts the stable trim point from 4° to 20°.

**Synergistic morphing increases the velocity range and reduces thrust.** We wanted to identify the drone's optimum morphology to achieve fast cruise flight and slow flight at a minimum thrust force. This implies to minimize drag while generating enough lift to stay aloft, respectively. Thus, we formulated a morphing shape optimization procedure (see Materials and Methods) to identify the three corresponding control inputs (elevator, tail sweep, symmetric wing sweep) that yield the minimum thrust force to achieve trim flight as a function of flight velocity.

At high flight velocities, we observe that the tucked wing and tail morphology is the preferred (Fig. 6A). At 12 m/s we show a thrust reduction by 59 % as compared to the extended wing and tail configuration. The LisHawk achieves a minimum thrust (0.37 N) at a cruise speed of 9.7 m/s. However, the tucked wing and tail morphology limits the minimum velocity to 7.6 m/s (Fig. 6A). When reducing the flight velocity below 7.6 m/s, we can show that the synergistic application of wing and tail is crucial: to counteract the drone's weight force at lower velocities, the wing must be extended (Fig. 6D) resulting in a nose up moment (Fig. 5B). This is balanced by a downward deflection of the extended tail (Fig. 6B and C), which generates additional lift. Beyond 20 degrees of angle of attack (Fig. 6E), however, we see that the extended tail has to deflect upward to maintain trim, which would reduce the overall lift and decrease the energy efficiency. Instead, the tail is tucked and deflected further downward (Fig. 6B), which adds to the overall lift. In doing so, we can show a notable reduction in flight velocity to 4.0 m/s (-44.5%) (Fig. 6A), with respect to the fully tucked configuration.

**Flight tests.** After a few preliminary flight tests to verify the LisHawk's airworthiness when morphing (Mov. S4), we proceeded to quantify the increase in aggressive flight behavior by example of a pull up maneuver (Fig. 7A). Here, the extended wing, tucked tail configuration was not considered as it is difficult to control due to its large pitch instability. We flew the LisHawk in trim flight with the tucked wing before a switch on the remote control was activated to deflect the elevator upward (-10°) (Mov. S5). Simultaneously**,** the drone adopted tucked wing and tail, tucked wing and extended tail, or extended wing and tail (Fig. 7B), while the drone control system logged the states of the drone (see Materials and Methods for the setup).

Our data indicate an increase in pitch acceleration by 445% for the extended wing and tail, as compared to the tucked wing and tail morphology (Fig. 7C). The increased acceleration in pitch leads to a temporary increase in angle of attack, which increases the lift and drag forces (Fig. 3) resulting in a linear acceleration by 383% (Fig. 7E). When comparing the flight trajectory, we see that the horizontal component is reduced from 9.5 m for the tucked wing and tail to 5.6 m for the extended wing and tail after 0.8 s (Fig. 7B). Similarly, the heading angle is increased from 10° to 68 °. This can be explained by the decreased linear velocity and the increased pitch velocity caused by the increased agility and maneuverability. We also see an improvement in agility and maneuverability with respect to the previously calculated values from our wind tunnel test results (Fig. 3 and 4), which could be explained by dynamic effects that are known to increase the aerodynamic forces and moments (*27*, *47*).

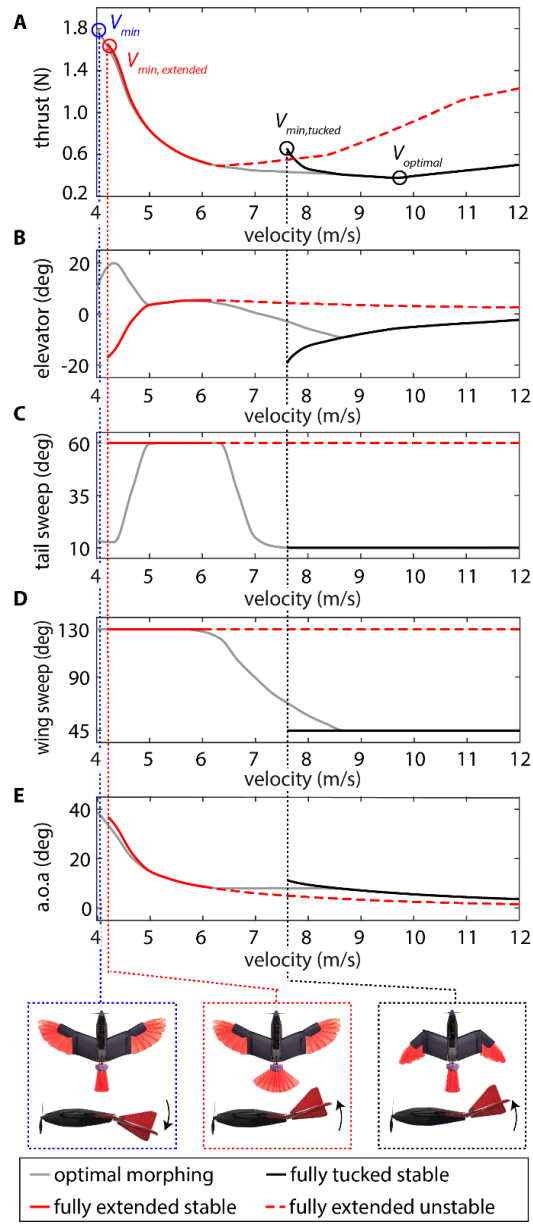

Fig. 6. **The optimal interplay of wing and tail morphing reduces the minimum velocity and the thrust force**. Graphs shown here are the results of an optimization study to minimize the thrust force with respect to the flight velocity by morphing (see Materials and Methods section for the setup). V indicates the flight velocity. (**A**) The fully tucked configuration reduces the thrust force at high velocities, while extending the wing and tail reduces the minimum flight velocity. (**B**) At low velocities, the elevator deflects downward, thus increasing the overall lift. (**C**) At low velocities, the tail shifts from extended to nearly tucked allow further increasing the angle of attack. (**D**) At velocities below 6 m/s the wing is fully extended, while at velocities over 8.5 m/s the wing is fully tucked. (**E**) To reach the minimum velocity of 4 m/s, the LisHawk operates in the high angle of attack (a.o.a) regime.

The tucked wing with the extended tail shows inferior performance during the pull up maneuver than the extended wing and tail configuration except for the pitch acceleration (Fig. 7C). This behavior could be linked to its reduced wing area and its high inherent stability (Fig. 5D). On one hand, the tucked wing produces less lift and drag than the extended wing (Fig. 3B), which reduces the linear acceleration (Fig. 7E) and decreases the change in flight velocity (Fig. 7F). On the other hand, the increase in angle of attack produced by the fast change in pitch (Fig. 7D), which may be in conflict with the elevator's limited control authority above 17° angle of attack (Fig. 4C). However, we think that the extended tail in combination with the tucked wing may be superior when performing aggressive maneuvers in fast flight to change the flight path at a lower cost in drag as compared to the fully extended configuration.

**Discussion**

In this study we asked how the synergistic application of wing and tail morphing inspired by the gliding northern goshawk could improve flight capabilities of winged air vehicles. We developed a novel morphing drone able of changing the sweep and the area of both wing and tail (Fig. 2). In doing so, we extend previous bio-inspired morphing designs (*9*, *27*) by a variable sweep and area tail, which is especially relevant for high maneuverability and agility (*18*, *48*). On the basis of tunnel measurements, shape optimization studies, and outdoor flight tests with the morphing LisHawk, we characterized the aerodynamic changes when morphing. The data showed that synergistic morphing of wing and tail permits large changes in maneuverability, agility, inherent stability and energy efficiency to increase performance in both aggressive and cruise flight. Here, we want to highlight our main findings.

First, to increase pitch agility, we found that the best morphology to produce large pitch moments is dependent on the angle of attack (Fig. 4D). At low angles of attack, large nose up pitch moments are produced when deflecting the extended tail upward (Fig. 4A). At high angles of attack, however, large pitch moments are predominantly produced by extending the wing, while the tail is tucked (Fig. 4B). Through the targeted application of wing and tail morphing, we can show that a large moment in pitch can be produced even at high angles of attack. This finding is especially relevant for perching maneuvers, during which a huge loss in airspeed and kinetic energy is desired for precision landings in confined spaces (*46*). The tremendous nose up pitch moment produced from synergistic morphing (Fig. 7F) induces large drag forces (Fig. 3A to C), which favor large decelerations (Fig. 7D and E). Instead, conventional aircraft tend to lose control power at high angles of attack, which limits aggressive flight behavior (*49*).

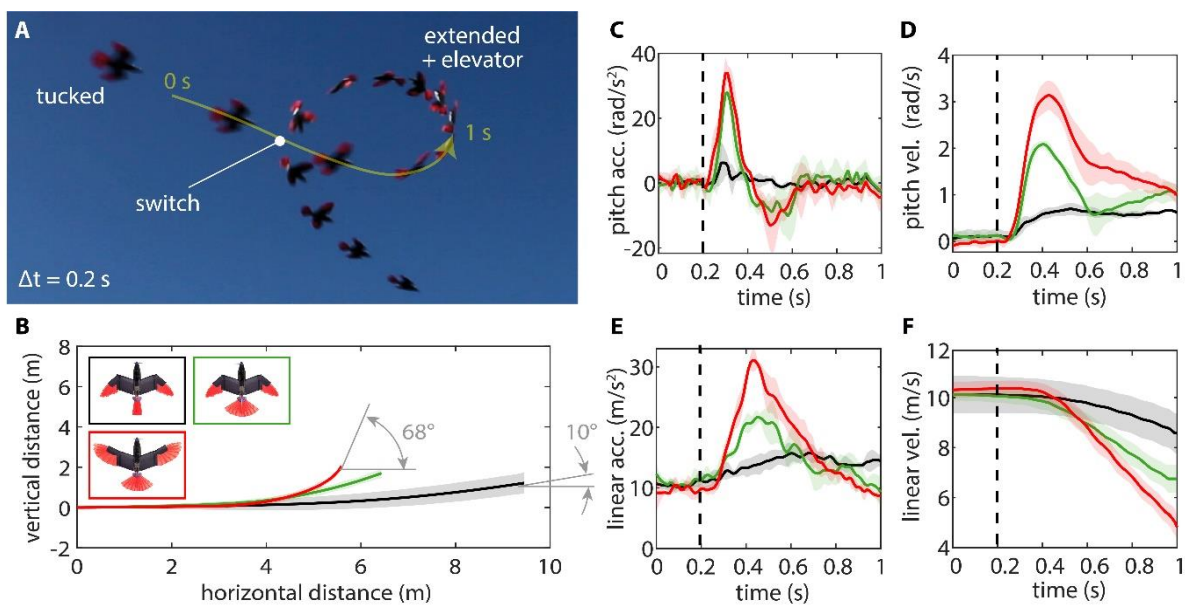

Fig. 7. **Synergistic morphing increases flight performance during pull up maneuver.** The shaded areas indicate the standard deviation from four trial runs. (**A**) Exemplary illustration of the pull up maneuver when fully extending wing and tail. (**B** to **F**) The LisHawk's response to an upward elevator deflection (-10°) and change in morphology to tucked wing and tail (black), tucked wing, extended tail (green), and extended wing and tail (red) (Mov. S5), with respect to the flight trajectory (**B**), the linear acceleration (and thus maneuverability) (**C**), the flight velocity (**D**), the pitch acceleration (and thus pitch agility) (**E**), and the pitch velocity (**F**).

Second, we found that our results are in line with previous theoretical studies on bird flight, which suggest that the role of the avian tail in combination with its morphing wing is not limited to the pitch control and to trim (*16*). Rather, it is argued that the tail also increases lift and reduces induced drag during high angles of attack flight (*13*, *16*, *42*). Our data support the hypothesis that additional lift is produced when applying the morphing tail in conjunction with the extended wing for increased maneuverability and during slow flight (Fig. 3C and Fig. 5D). We could show a downward deflection of the tail to balance the nose up moment when extending the wing at low flight velocities (Fig. 6B). This increases the overall camber (curvature of the airfoil) of the wing-tail-entity, which allows to fly slower (Fig. 6A) and to reduce induced drag. In doing so, the minimum flight velocity could be lowered from 7.6 m/s with the tucked wing and tail to 4.0 m/s with the extended wing and tucked tail.

Third, our data showed that, like birds (*13*), we could improve the flight performance by changing the inherent pitch stability as a result of synergistic morphing of wing and tail. While changing the wing's sweep angle is known to alter the inherent pitch stability (*27*, *29*), no previous study has experimentally studied the impact of a variable sweep and area tail on drones. We show that the extended tail increases pitch stability while maintaining the same trim point as the fully tucked configuration (Fig. 5B). While this insignificantly affects drag at low angles of attack (Fig. 3D), we hypothesize that this could be efficient mean of rejecting disturbances as opposed to an active tail deflection. Our data also suggests that the northern goshawk-like airframe geometry causes high angle of attack trim points (Fig. 5D). This is especially relevant as it suggests that synergistic morphing of wing and tail permits controlled and stable flight in the high angles of attack regime, which is also known as supermaneuvrability (*50*). This is a distinctive feature of highly agile and maneuverable aircraft.

Fourth, we could show that synergistic morphing of wing and tail benefit fast cruise flight by reducing drag and thus the required thrust by 59 % at a flight velocity of 12 m/s when switching from the extended wing and tail to the tucked wing and tail (Fig. 6A, C and D). This reduction can be explained by the low exposed wing and tail area which reduces parasitic drag (*41*). Furthermore, tucking both wing and tail makes the drone inherently stable in pitch and reduces the wing area, which could decrease the drone's sensitivity to wind gusts (*9*, *11*).

Fifth, we extended previous research on asymmetric morphing for roll control (*9*, *30*). Our data suggests that asymmetric sweeping can produce large roll moments at high angles of attack (Fig. 4F). This is in contrast with ailerons used on conventional aircraft that are prone to a large roll moment reduction or even roll moment reversal at high angles of attack, which limits agility (*29*, *51*). Indeed, we can conclude that asymmetric sweeping is superior to ailerons during aggressive flight.

Although our flight tests showed a clear trend towards increased mission capability, the full potential of the morphing surfaces applied on the LisHawk could not be captured due to the human operator in the loop. On one hand, we actuated the throttle, asymmetric sweep for roll, elevator, and rudder in a continuous manner, while the symmetric sweep of wing and tail were limited to three configurations (tucked, intermediate, and extended). On the other hand, we were restricted to operate in the stable flight regime during our flight tests. To show the full potential of synergistic morphing of wing and tail it is necessary to perform tests with an autopilot in the loop which could take advantage of the unstable flight regime and the continuous adaption of all degrees of freedom simultaneously (*28*, *41*, *52*). Furthermore, wind tunnel tests performed for this study provided the static aerodynamic and flight mechanic properties of the LisHawk drone. However, rapid changes in angle of attack during dynamic maneuvers delay flow detachment to higher angles of attack and can cause a considerable increase of lift force during a short time period (*27*, *47*). The endeavor to obtain such measurements is challenging, yet they could further reinforce the case and broaden the understanding of avian-inspired synergistic morphing on flight performance.

We think that synergistic morphing to match avian gliding flight as applied on the LisHawk drone could facilitate extended intelligence, surveillance and reconnaissance (ISR) missions in growing urban environments. The drone's low parasitic drag and its reduced sensitivity to wind when tucking its wing and tail could extended the mission range with reduced energy expenditure. Vast areas could be covered in little time. When a specific target is identified, a synergistic morphing drone could acquire data at a closer proximity while negotiating through obstacle-cluttered and confined airspaces due to its aggressive and slow flight qualities. With its extended lifting surfaces, the morphing drone could induce high accelerations even at low velocities. ISR missions could be further enhanced by placing sensors at points of interest, by resting on higher ground for long term observations, or by making stops for recharging (*1*, *53–55*). To do so, we believe that an avian-inspired morphing wing and tail drone can address the fundamental challenge of precision landings through its rapid deceleration capabilities that minimize the landing footprint. Perching maneuvers supported by high angles of attack stable flight capabilities could permit reliable landings on ledges or on power lines (*56*). Overall, the avian-inspired synergy of morphing wing and tail could notably increase the scope of applications for fixed-wing drones.

## Materials and Methods

**Fabrication of the feathers, morphing surfaces and fuselage.** For the morphing wing and tail (see Supplementary Material for the airfoil selection), we manufactured artificial feathers that are comprised of three major parts (Fig. 2E): First, we cut a skeleton from a 0.3 mm fiber glass sheet with a CO2 laser cutter (Trotec Speedy 400). This material was chosen because it combines a low mass, flexibility, and sufficient stiffness to absorb the aerodynamic loads when the feathers are slightly overlapped. Second, we covered the skeleton with an airtight and tear-resistant membrane made ripstop polyester fabric (Icarex) with cyanoacrylate glue (viscosity: 3-10mPas). Last, we fixed a 1.5 mm carbon tube onto the skeleton to act as a shaft.

The fabrication process of the morphing wing mechanism (Fig. 2C) is depicted in Mov. S1. We used 3D-printed (Stratasys EDreamer) acrylonitrile butadiene styrene (ABS) part that is reinforced by two 0.5 mm glass fiber plates (cut by CO2 laser cutter) laser, and an aluminum ring. For the wrist skeleton, we chose a balsa/glass fiber (thickness 6 and 0.7 mm) composite structure, which entails both a low weight and high rigidity. To further reinforce the wrist skeleton, wrapped the most stressed parts with Kevlar (0.2 mm, 20 kg). We connected the bearings ($D = 7$ mm, $d = 3$ mm, $W = 3$ mm) with the glass fiber parts by using custom made ABS rings which were fixed with cyanoacrylate glue (viscosity: 3-10mPas). The feathers are attached to feather pins, which in return are attached to the skeleton via 2 mm carbon shafts. Inviscid cyanoacrylate glue is used to interconnect parts.

The fabrication process of the morphing and deflecting tail shown in Fig. 2G, consists of the steps shown in Mov. S2. For the tail skeleton, we manufactured interlocking elements from a 0.5mm glass fiber (cut with the CO2 laser cutter), which were fixed with cyanoacrylate glue. In doing so we could nearly halve the weight with respect to the previous version, which was fully 3D-printed in ABS. However, we made the pulleys, the feather pins and the universal joint from 3D-printed ABS, while we chose 2m carbon pipes for all the pins. A Kevlar cable (0.2 mm, 20 kg) is used to connect the outer feather pin levers over the skeleton pulleys, via fuselage beam to the servo at the front of the fuselage.

We attached the morphing wing (via t-connectors) and tail (via universal joint) to a 8x8mm$^2$ (thickness: 0.5 mm) pultruded carbon fiber square tube in the center of the fuselage (Fig. 2A), which acts as a backbone that provides stiffening for the aerodynamic body and absorbs all aerodynamic forces generated by the lifting surfaces. It is surrounded by an EPP shell with a thickness of 6 mm, which we cut around wooden patterns with a hot wire. To achieve the rounded shape, we separated the fuselage into three parts (front, center, back) which are united with UHU Por and the surface was smoothened with a hot iron. The finished shell hides the structural and electrical components in an aerodynamically favorable manner. We mounted the motor on the LisHawk's EPP fuselage and did now link it directly to the square tube (Fig. 2A). This way we could dampen vibrations from the motor and make the drone more crash resilient.

**Drone control system.** For the propulsion system (Fig. 2A), we chose a S-1805-2250 KV brushless DC motor with a 7x6 GWS propeller (nominal stationary thrust: $F_T = 257$ g, battery: 2S 450 mAh) and a Turnigy Plush 10 A electronic speed controller. To actuate the wrist flexing mechanism, we chose two KST X08-Plus servos ($m = 9$ g, $T = 3.8$ kgcm @ 6V), while for the tail deflections as well as the tail spreading we used three BMS-306 digital servos ($m = 7.1$ g, $T = 2.0$ kgcm @ 6V). We used a PixRacer for data logging in combination a Drotek M8Q GPS block with antenna. We piloted the LisHawk drone with a Turnigy X9R PRO 2.4GHz radio controller and an Orange RX110 DSMX/DSM2 satellite receiver.

**Wind tunnel measurement setup.** We performed wind tunnel studies on the open-jet wind tunnel (from WindShape) at the haute école du paysage, d'Ingenierie et d'architecture in Geneva. We set the air stream to the expected mean velocity of 8 m/s, which corresponds to a Reynolds number of 91'837 for the LisHawk drone. We mounted the LisHawk drone in its center of gravity on a RUAG Aerospace 6 component sting balance, which was attached to a robotic arm to accurately and autonomously position the drone in the wind tunnel at the respective angle of attack. For data logging, we used a HBM MX 840B universal amplifier and the Catman V5.2.2 software for data acquisition. Before a measurement sequence was started, we zeroed the data logging device in calm air conditions. Then, we measured the aerodynamic forces and moments for 4 seconds at 300Hz after a setting time of 12 s. We measured from an angle of attack of -4° to 34° at 2° steps. To analyze the data, we used a custom made MATLAB script. We used the tucked wing, tucked tail configuration (S = 0.117m$^2$, mean aerodynamic chord = 0.161 m) as a baseline to calculate the aerodynamic coefficients. The wind tunnels turbulence was 0.5 %, the deviation of flow velocities in the test section was < 5 % and the error for the angle of attack was < 0.1 °. Aerodynamic coefficients were calculated with the standard definitions described in (*5*) (Supplementary Material). All measurement points used in this manuscript and their corresponding standard deviation can be found in the supplementary material (Fig. S4 and Fig. S5).

**Maneuverability metrics.** Maneuverability is a concept of linear motion. We define it as the ability to induce a high controlled linear accelerations, $\ddot{x}$ in $x$-direction, $\ddot{y}$ in $y$-direction, and $\ddot{z}$ in z-direction (Fig. S2), to rapidly change the linear velocity and direction of the flying body's translational movement (*7*). In the wind fixed frame, it can be defined as (Supplementary Material for mathematical formulation and assumptions)

$$\begin{pmatrix} \ddot{x} \\ \ddot{y} \\ \ddot{z} \end{pmatrix} = \begin{pmatrix} \frac{-D}{m} - g \sin\alpha \\ -\frac{L \sin\varphi}{m} \\ \frac{L \cos\varphi}{m} - g \cos\alpha \end{pmatrix}, \qquad (1)$$

with $D$ being the drag force, $L$ being lift force, $m$ being the mass, $g$ being the gravitational constant, $\alpha$ being the angle of attack, and $\varphi$ being the bank angle. Consequently, to increase maneuverability aerodynamic forces $D$ and $L$ must be increased, while the aircrafts mass $m$ should be decreased.

**Agility metrics.** Agility is a concept based on rotational motion. We define it as the ability produce a high controlled angular rate in roll $\dot{p}$, pitch $\dot{q}$, and yaw $\dot{r}$, which are also called heading accelerations that act around the body fixed axes (Fig. S2). To estimate external factors influencing agility, the kinematic equation of motion is simplified (see Supplementary Material for mathematical formulation and assumptions) to pure rotation (*57*), so that

$$\begin{pmatrix} \dot{p}_{pure} \\ \dot{q}_{pure} \\ \dot{r}_{pure} \end{pmatrix} = \begin{pmatrix} \frac{P}{I_x} \\ \frac{Q}{I_y} \\ \frac{R}{I_z} \end{pmatrix}, \qquad (2)$$

with $P$ being the roll moment, $Q$ being the pitch moment, and $R$ being the yawing moment acting around the aircraft fixed axes. The denominator constitutes of the moment of inertia $I_x$ around $x$, $I_y$ around $y$, and $I_z$ around $z$. Thus, to increase agility we must increase the aerodynamic moments $P$, $Q$, and $R$ or decrease the moments of inertia in the respective body fixed $I_x$, $I_y$, and $I_z$ axes.

**Longitudinal static stability.** Mathematically, an aircraft is stable in pitch if the derivative of the pitch coefficient $C_m$ with respect to the angle of attack (Fig. 4E) also called the pitch stiffness (8), is negative such that

$$C_{m,\alpha} = \frac{\partial C_m}{\partial \alpha} < 0. \qquad (3)$$

Thus, if this applies, the aircraft exhibits positive static stability as the angle of attack will passively converge towards a stable equilibrium at $C_m = 0$ following a disturbance. Contrarily, if the pitch stiffness is positive so that $C_{m,\alpha} > 0$, the angle of attack will diverge from equilibrium and the aircraft is negative static stable. The independent contribution of wing and tail are provided in the Supplementary Material.

**Modelling assumptions for the shape optimization.** In search of the global minimum thrust, we solved the optimization problem using MATLAB's *multistart* algorithm with the *fmincon* solver (*46*) (see Supplementary Material). This algorithm selects the control inputs with the overall minimum thrust based on multiple searches of the local minima from an extensive range of initial points. We developed the optimization framework based on the aircraft's longitudinal body axes frame using the wind tunnel's aerodynamic data without propeller slipstream effects. The thrust control input represents the cost function which we minimized. To ensure steady straight level flight control solutions, we specified zero-equality constraints on the aircraft's pitch moment, pitch rate, flight path angle and flight acceleration.

**Flight test setup.** We launched LisHawk by arm throw and flights with durations between 4 and 8 minutes were performed before landing on the ground (Mov. S4). As continuous control inputs, throttle for thrust, elevator for pitch, rudder for yaw and asymmetric wing sweep for roll were available. We defined switches on the remote control to change the sweep of the morphing tail and main wing symmetrically during the flight such that $\delta_t = 10\,°$ and $60\,°$ and $\delta_w = 45\,°$, $90\,°$, and $130\,°$, respectively (Fig. 2A). For the pull up study, we initiated the maneuver from trim with the tucked wing and at a 60% thrust setting. Through a switch on the remote control, we initiated the upward tail deflection (-10°), while remaining tucked, extending the tail, and extending both wing and tail. We logged the linear acceleration, the rotational velocity, the control inputs, and the extended Kalman Filter estimate of the position/velocity with the PixRacer (sampling rate PixRacer: 100Hz, Sampling rate GPS: 10 Hz). We then used a custom made MATLAB script to calculate the mean values over time and their standard deviations. We aligned the horizontal and vertical flight trace when the elevator deflection was engaged and rotated each trial run to fly from left to right, as shown in Fig. 7B. We also aligned the linear acceleration, rotational acceleration, linear velocity and rotational velocity when the elevator deflection was engaged (Fig. 7B-F). Due to GPS tracking errors, we had to exclude one trial from each run (five were done for each configuration). No autopilot or other autonomous flight enhancing measures were implemented during flight. All flights took place on a large, open field in calm wind conditions (measured wind speeds below 2 m/s at 3.5 m above ground).

## Supplementary Material

*https://zenodo.org/record/3648670*
Supplementary Text
Fig. S1. Airfoil selection.
Fig. S2. Coordinate systems and variables.
Fig. S3. Wing and tail contribution to pitch stability.
Fig. S4. Standard deviation of lift and drag coefficient plots.
Fig. S5. Standard deviation of pitch and roll coefficient plots.
Mov. S1. Morphing wing fabrication.
Mov. S2. Morphing tail fabrication.
Mov. S3. Morphing control surfaces.
Mov. S4. Flight test.
Mov. S5. Pull up maneuver
References (*5*, *8*, *10*, *17*, *46*, *58*)

**Acknowledgements:** We thank Guillaume Catry, Nicolas Bosson, Sergio Marquez, and Alberic Gros from WindShape for letting us use their innovative wind tunnel facility and for their continuous support. We would also like to thank Vivek Ramachandran and Olexandr Gudozhnik for their help during the writing process and their technical support, respectively.

**Funding:** This study was supported by the Swiss National Science Foundation through the National Center of Competence in Research Robotics.

**Author contributions:** E.A., M.F., S.M. and D.F. developed the concept and wrote the manuscript. E.A. designed and fabricated the LisHawk drone. E.A., M.F. and F.N. designed the experiment and performed experimental wind tunnel tests. M.F. developed the test environment for the shape optimization study and performed simulations. E.A. conducted and analyzed the flight tests.

**Competing interests:** There are no competing interests.

**Data and materials availability:** E.A. may be contacted for additional information.